\documentclass{aa}
\usepackage{psfig}
\def\la{\;
\raise0.3ex\hbox{$<$\kern-0.75em\raise-1.1ex\hbox{$\sim$}}\; }
\def\ga{\;
\raise0.3ex\hbox{$>$\kern-0.75em\raise-1.1ex\hbox{$\sim$}}\; }

\begin{document}

\title{Metal abundances and kinematics of quasar absorbers}

\subtitle{I. Absorption systems toward J2233--606\thanks{Based 
on public data released from
UVES Commissioning at the VLT Kueyen telescope (ESO, Paranal,
Chile), the New Technology Telescope (ESO, La Silla, Chile) 
and the NASA/ESA Hubble Space Telescope obtained
at the Space Telescope Science Institute, which is operated by
the Association of Universities of Research in Astronomy, Inc.
under NASA contract NAS 5-26555.}
}

\author{
S. A. Levshakov\inst{1}\thanks{On leave from the
Ioffe Physico-Technical Institute, Russian Academy of Sciences,
St.~Petersburg}
\and I. I. Agafonova\inst{2}
\and M. Centuri\'on\inst{3}
\and I. E. Mazets\inst{2}
}

\offprints{S. A. Levshakov}

\institute{
Division of Theoretical Astrophysics,
National Astronomical Observatory, Mitaka, Tokyo 181-8588, Japan
\and
Department of Theoretical Astrophysics,
Ioffe Physico-Technical Institute, 
194021 St. Petersburg, Russia
\and
Osservatorio Astronomico di Trieste, Via G. B. Tiepolo 11,
34131 Trieste, Italy\\
}

\date{Received 00  / Accepted 00 }

\abstract{
The metal line profiles of different ions observed in 
high \ion{H}{i} column density
systems [$N$(\ion{H}{i}) $> 10^{16}$ cm$^{-2}$] 
in quasar spectra can be used to constrain the
ionization structure and kinematic characteristics of the absorbers.
For these purposes, a modified Monte Carlo Inversion (MCI) procedure
was applied to the study of three absorption systems in the spectrum of 
the HDF-South quasar J2233--606 obtained with
the UVES spectrograph at the VLT/Kueyen telescope. The MCI does not confirm
variations of metal abundances within  separate systems which were discussed
in the literature. Instead, we found that an assumption of a homogeneous
metal content and a unique photoionizing background is sufficient to describe
the observed complex metal profiles. 
It was also found that the
linear size $L$ and the line-of-sight velocity dispersion $\sigma_{\rm v}$
measured within the absorbers obey a scaling relation, namely, 
$\sigma_{\rm v}$ increases with increasing $L$, and that
metal abundance is inversely proportional to
the linear size of the system: the highest metallicity was measured in 
the system with the smallest $L$.
}

\maketitle

\section{Introduction}
 
\addtocounter{footnote}{3}

Absorption systems in quasar spectra provide unique information
on the intervening intergalactic matter (IGM) up to redshift
$z \simeq 6$, back to the time when the Universe was less than 7\%
of its present age.
High resolution spectroscopic observations available nowadays 
at large telescopes open new
opportunities to investigate the physical nature of quasar absorbers. 
Reliable data on the
chemical composition of the IGM and on the
physical characteristics
(like velocity and density distributions, 
volumetric gas density, kinetic temperature,
ionization structure etc.)  of the
absorbers is an important 
clue to our understanding of galaxy formation, chemical evolution
of the IGM
and the origin of the large-scale structure.

In recent investigations much attention find the metal
systems  which are the absorbers exhibiting 
as a rule numerous lines of low (like 
\ion{H}{i}, \ion{C}{ii}, \ion{Si}{ii},  \ion{Mg}{ii}, 
\ion{Al}{ii}) 
and high (like 
\ion{C}{iii}, \ion{N}{iii}, \ion{Si}{iii},
\ion{C}{iv}, \ion{Si}{iv}, \ion{N}{v}) ionized species.

Presence of metals provides a unique opportunity to study
the physical conditions of matter at early epochs.
Unfortunately, the computational 
methods usually applied to high resolution spectra 
lie quite often  behind the quality of observational
data and fail to extract from them 
all encoded information. 
The common processing method consists of the deconvolution 
of complex absorption 
profiles into an arbitrary number of separate
components (assuming a constant gas density within each of them)
which are then fitted to Voigt profiles. 
However,
in many cases this procedure may not correspond to 
real physical conditions: observed complexity 
and non-similarity of the profile shapes of different ions
indicate that these systems are in general absorbers 
with highly fluctuating density and velocity fields
tightly correlated with each other.
Too high or too low gas temperatures, extremely varying metallicities
between subcomponents, exotic UV background spectra and other
physically badly founded outcomings may be artifacts arising 
from the inconsistency of the applied methods
(see examples in
Levshakov et al. 1999;  
Levshakov et al. 2000b, hereafter Paper~I).

In Paper~I we developed a new method for the QSO spectra 
inversion, -- the Monte Carlo Inversion (MCI), --
assuming that the absorbing region is a cloud 
with uniform metallicity but with fluctuating density and
velocity fields inside it. 
This computational procedure 
which is based on stochastic optimization 
allows us to recover both the underlying hydrodynamical fields and 
the physical parameters of the gas.
First application of the MCI to the analysis of the
$z_{\rm abs} = 3.514$ system toward Q08279+5255
(Levshakov et al. 2000a) 
has shown that the proposed method is very
promising especially
in the inversion of complex absorption spectra 
with many metal lines. 

In this paper we
start a new comprehensive 
survey of the metal systems for which high resolution and
high signal-to-noise spectra are available. 
We present here the results for three absorption systems 
($z_{\rm abs}$ = 1.87, 1.92 and 1.94) from the spectrum 
of the quasar J2233--606 which have been already studied
by Prochaska \& Burles (1999), and 
D'Odorico \& Petitjean (2001, hereafter DP) 
using the common Voigt fitting method. 
We re-calculate
these systems using the MCI in order 
to compare the applicability of both approaches 
and to show up their restrictions.

The structure of the paper is as follows. 
In Sect.~2 the data sets used in the MCI analysis are described.
Sect.~3 contains the details of the applied 
computational procedure. The results obtained for each of the
mentioned above systems are presented in Sect.~4. 
Conclusions are reported in
Sect.~5. 
In Appendix the general equation of the entropy production
rate is given which is used to calculate
the density and velocity configurations
along the line of sight exhibiting minimum dissipation.

\begin{figure*}
\vspace{0.0cm}
\hspace{0.0cm}\psfig{figure=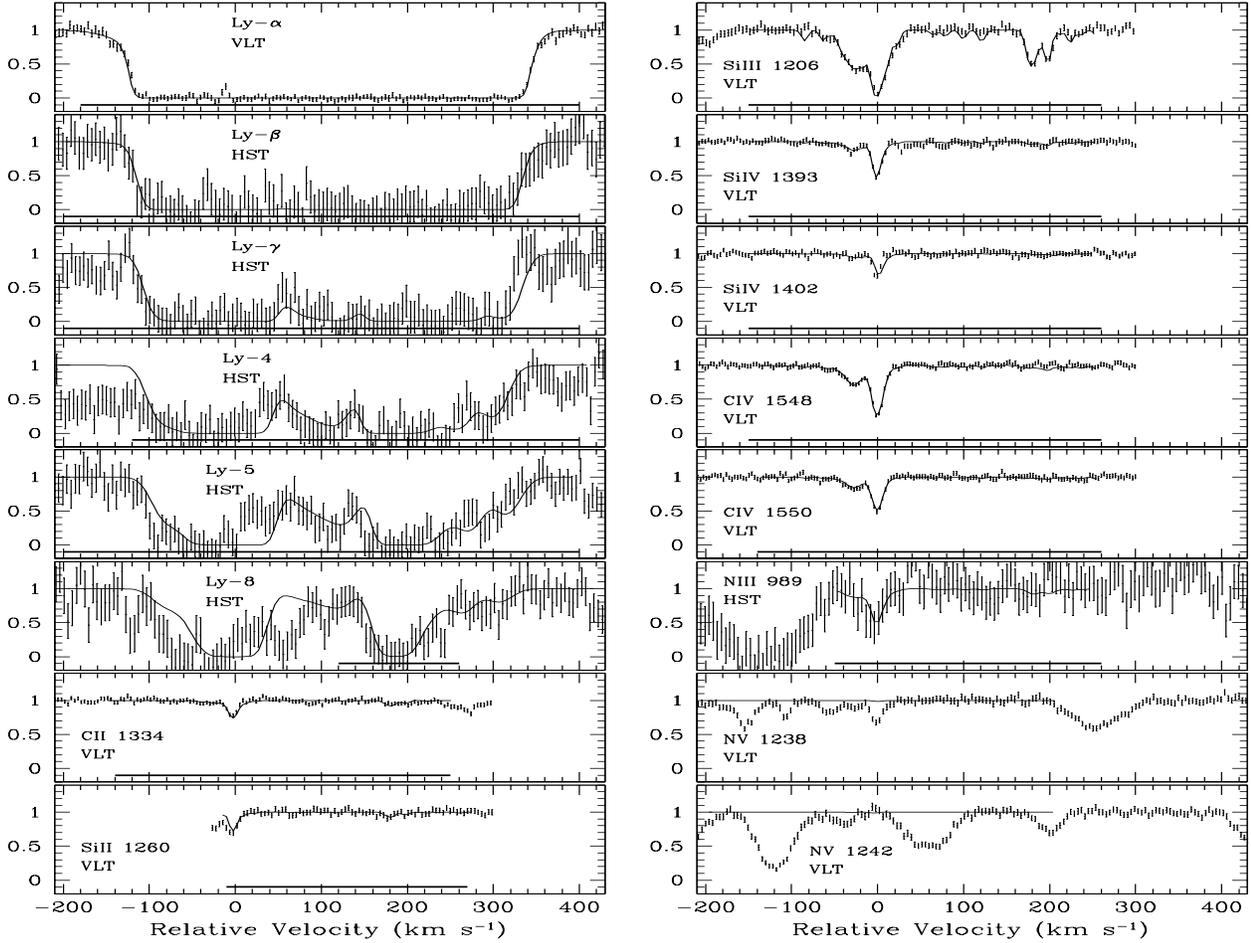,height=13.0cm,width=20.0cm}
\vspace{-0.5cm}
\caption[]{
Hydrogen and metal absorption lines associated with the
$z_{\rm abs} \simeq 1.92$ system toward J2233--606
(normalized intensities are shown by dots with $1\sigma$ error bars).
The zero radial velocity is fixed at $z = 1.92595$. Smooth lines are
the synthetic spectra convolved with the corresponding point-spread
functions (FWHM$_{\rm VLT} = 6.7$ km~s$^{-1}$, 
FWHM$_{\rm HST} = 10.0$ km~s$^{-1}$) and computed with the physical
parameters from Table~1. Bold horizontal lines mark pixels included
in the optimization procedure. 
The normalized $\chi^2_{\rm min} = 1.10$ (the number of degrees of
freedom $\nu = 1450$) 
}
\label{fig1}
\end{figure*}

\section{Data}

High-quality data of the Hubble Deep Field South (HDF-S) quasar
J2233--606 ($z_{\rm em} = 2.238$, B $\simeq 17.5$) 
were obtained during the
commissioning of the UVES on the VLT 8.2m Kueyen telescope 
at Paranal (Chile) in October 1999.   
The resolving powers at which the spectra were recorded 
in the spectral ranges $\lambda\lambda = 3050-10000$ \AA\
(J2233--606) and 
$\lambda\lambda = 3700-5000, 6670-9100$ \AA\ (Q0000-2620) 
were
$R \simeq 45000$ and $R \simeq 49000$,
which correspond to velocity resolutions of
FWHM $\simeq 6.7$ km~s$^{-1}$ and $\simeq 6.1$ km~s$^{-1}$,
respectively. 

The data reduction and the identification of metal absorption-line
systems in the J2233--606 spectrum
are reported in Cristiani \& D'Odorico (2000). 
In our study we also used the J2233--606 echelle spectrum
($R \simeq 30000$, $\lambda\lambda~2275-3118$ \AA) 
obtained with the HST/STIS 
(Savaglio 1998).

\section{Computational procedure}

The complete description of our computational procedure
is given in Paper~I. Here we summarize briefly basic 
model assumptions
and emphasize new details recently included in the MCI. 

We assume that all lines observed in a metal system arise 
in a continuous absorbing gas slab of a thickness $L$ (presumably
the outer region of a foreground distant galaxy).
The absorber exhibits a fluctuating gas density and a mixture 
of bulk motions such as infall and
outflows, tidal flows etc.,
resulting in a stochastic velocity field.
Metal abundances are assumed to be constant within the absorber 
and gas is supposed 
to be optically thin for the ionizing UV radiation. 

Within the absorbing region the radial velocity  $v(s)$ and 
total hydrogen density $n_{\rm H}(s)$ distributions 
along the line of sight are 
the same for all ions. In the computational procedure these
two random fields
are represented by their sampled values
at equally spaced intervals $\Delta x$ 
($x$ is dimensionless radial coordinate $s/L$), 
i.e. by the vectors
$\{ v_1, \ldots, v_k \}$ and $\{ n_1, \ldots, n_k \}$ 
with $k$ large enough
to describe the narrowest components of complex spectral lines. 

Further we suppose the thermodynamic and ionization equilibrium at
each computational point along the sightline which means that
fractional ionizations of different ions are determined 
exclusively by the gas density and vary from point to point.
These fractional ionization variations are just the cause 
of the observed 
diversity of profile shapes. 
To calculate the kinetic temperature and fractional 
ionization of ions the photoionization code 
CLOUDY (Ferland 1997) was used. 

The inputs  to CLOUDY are the dimensionless ionization parameter 
$U = n_{\rm ph}/n_{\rm H}$ ($n_{\rm ph}$ -- the number density 
of photons with energies above 1 Ry), 
metallicity and the background ionizing spectrum for which  
the Haardt-Madau spectrum (HM) was adopted (Haardt \& Madau 1996).
The number density of the ionizing photons for this spectrum
\begin{equation}
n_{\rm ph} = \frac{4\pi}{c\,h}J_{912}\,\int^\infty_{\nu_{\rm c}}\,
\left(\frac{J_\nu}{J_{912}}\right) \frac{d\nu}{\nu}\; ,
\label{eq:E1}
\end{equation}
equals at $z = 2$ to 
$2.26\times10^{-5}$ cm$^{-3}$ 
[here $c$, $h$, $\nu_{\rm c}$ and $J_\nu$ are
the speed of light, the Planck constant, the
frequency of the hydrogen Lyman edge, and the specific intensity (in
ergs cm$^{-2}$ s$^{-1}$ sr$^{-1}$ Hz$^{-1}$) at frequency $\nu$, respectively].

Fractional ionization curves $\Upsilon (U)$  
were computed with CLOUDY for solar abundance pattern and different 
metallicities  and then included  
in the MCI code to calculate the optical depths for ions involved in 
the fitting.
If the obtained solution revealed the abundance pattern different from the solar
one, $\Upsilon (U)$ were recalculated for this new pattern and all computations
repeated. It should be noted, however, that differences of $0.2\div0.3$ dex
from solar values influence the fractional ionizations only weakly. 

The values of velocity and density at subsequent computational points 
are considered to be correlated and are described  by means of 
Markovian processes. In particular, the velocity is computed as follows:
\begin{equation}
v(x + \Delta x) = f_{\rm v}\,v(x) + \epsilon(x + \Delta x)\; ,
\label{eq:E3}
\end{equation}
where $f_{\rm v} = R_{\rm v}(\Delta x)/\sigma^2_{\rm v}$, 
$R_{\rm v}$ being the correlation between
the velocity values at points separated by a distance $\Delta x$, i.e.
$R_{\rm v} = \langle v(x + \Delta x)\,v(x) \rangle$, 
$\sigma_{\rm v}$ the 
velocity dispersion of bulk motions, and $\epsilon$ a random normal
variable with zero mean and dispersion
\begin{equation}
\sigma_{\epsilon,{\rm v}} = \sigma_{\rm v}\sqrt{1 - f^2_{\rm v}}\; .
\label{eq:E4}
\end{equation}

\begin{figure}
\vspace{0.0cm}
\hspace{-0.15cm}\psfig{figure=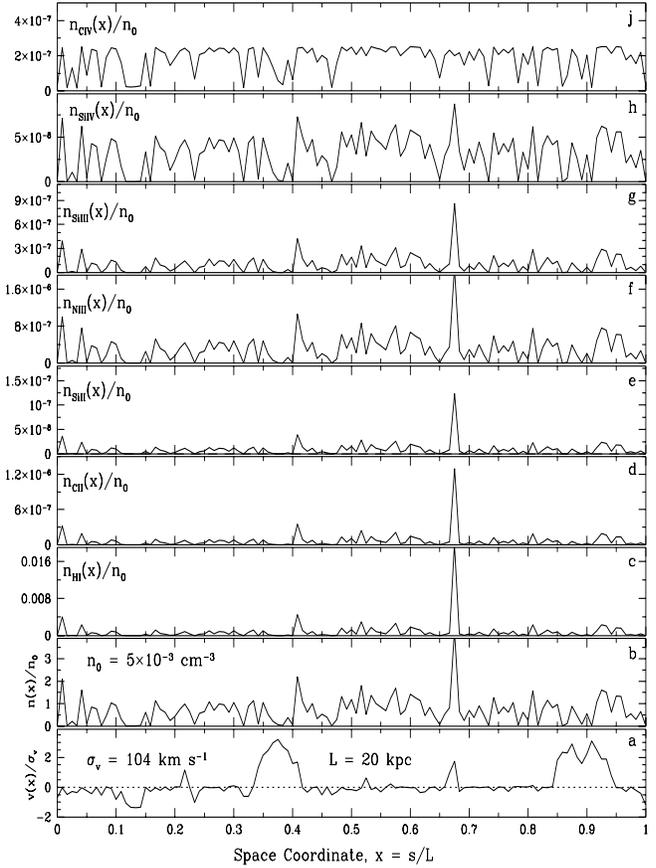,height=12.0cm,width=9.5cm}
\vspace{-0.5cm}
\caption[]{
Computed velocity (panel {\bf a}) 
and density of gas (panel {\bf b}) and ions
(panels {\bf c}--{\bf j}) 
distributions along the line of sight for the system at $z = 1.92595$
toward J2233--606. Shown are patterns rearranged according to the
principle of minimum entropy production rate (see text)
}
\label{fig2}
\end{figure}

\begin{figure}
\vspace{0.0cm}
\hspace{-0.15cm}\psfig{figure=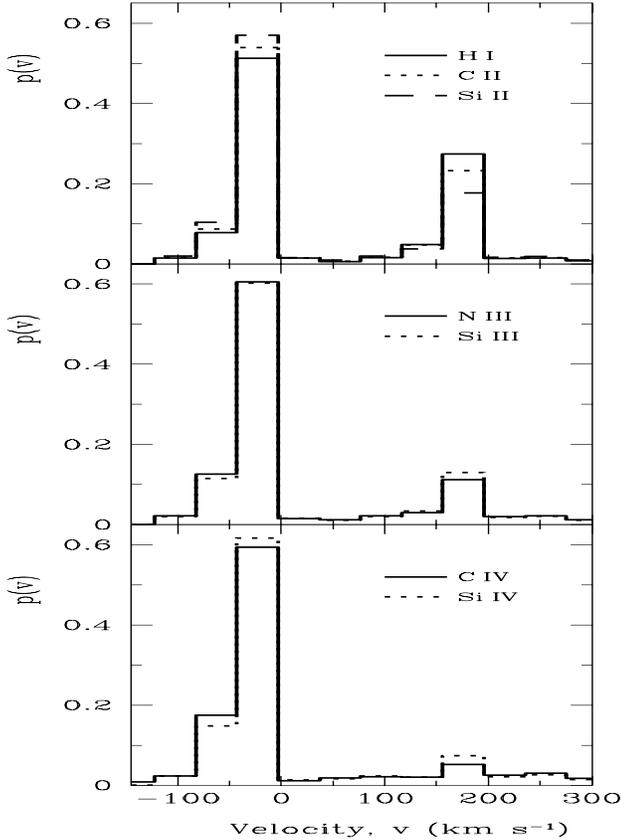,height=12.0cm,width=12.5cm}
\vspace{-0.5cm}
\caption[]{
Density-weighted velocity distribution functions, $p(v)$,
for \ion{H}{i}, \ion{C}{ii}, \ion{Si}{ii}, \ion{N}{iii},
\ion{Si}{iii}, \ion{C}{iv}, and \ion{Si}{iv} as restored 
by the MCI procedure in the $z = 1.92595$ system toward J2233--606 
}
\label{fig3}
\end{figure}

Density is supposed to have a log-normal 
distribution (characterized by the mean 
$n_0$ and the second central moment $\sigma_{\rm y}$ 
of dimensionless variable 
$y$ = $n_{\rm H}$/$n_0$ ) and is described with help of 
an auxiliary Markovian process $\nu (x)$:
\begin{equation}
\nu(x + \Delta x) = f_{\nu}\,\nu(x) + \epsilon(x + \Delta x)\; ,
\label{eq:E5}
\end{equation}
where  $f_{\nu} = R_\nu(\Delta x)/\sigma^2_\nu$,  
$R_\nu$ being the correlation between
the values of $\nu$ at points separated by a distance $\Delta x$,
$\sigma_\nu$ the logarithmic density 
dispersion, 
$\sigma_\nu = \sqrt{\ln(1 + \sigma^2_{\rm y})}$\,  
and $\epsilon$ a random normal
variable with zero mean and dispersion 
$\sigma_{\epsilon,\nu} = \sigma_\nu\sqrt{1 - f^2_{\nu}}$\ .

Having defined $\nu(x)$, 
the total hydrogen density can be obtained as
\begin{equation}
n_{\rm H}(x) = n_0\,{\rm e}^{\nu(x) - \frac{1}{2}\sigma^2_\nu}\: .
\label{eq:E6}
\end{equation}

The optical depth $\tau_{{\rm a},i}(\lambda)$ 
at wavelength $\lambda$ for element `a' in $i$th ionizing stage 
is calculated from the equation:
\begin{eqnarray}
\lefteqn{
\tau_{{\rm a},i}(\lambda) = 
k_0 Z_{\rm a} N_0\,{\rm e}^{-\frac{1}{2}\sigma^2_\nu} } \nonumber \\ 
& & \times \int^1_0\,
{\rm e}^{\nu(x)}\,\Upsilon_{{\rm a},i}[U(x)]\,
\Phi_\lambda[\Delta\lambda_{\rm D}(x),v(x)]\,{\rm d}x\; ,
\label{eq:E7}
\end{eqnarray}
where $k_0$ is a constant, 
$N_0 = n_0\,L$ -- the expectation value of the total hydrogen column 
density, $Z_{\rm a} = n_{\rm a}/n_{\rm H}$ -- 
the relative abundance of element `a', 
$\Upsilon_{{\rm a},i}[U(x)]$ -- the fractional ionization 
for stage $i$, 
$\Phi_\lambda$ -- the profile function, 
$\Delta\lambda_{\rm D}$ -- the Doppler width of the line.

Before being compared with the observed spectrum, 
the synthetic intensities $\exp[-\tau (\lambda )]$ 
are convolved with the spectrograph point-spread function.

Thus the proposed model is fully defined by specifying the 
following values: the velocity vector $\{v_i\}$, 
the total hydrogen density vector $\{n_i\}$, 
the total hydrogen column density $N_0$, 
the mean ionization parameter $U_0$, 
the radial velocity dispersion $\sigma_v$, 
the density second central moment $\sigma_{\rm y}$, 
the element abundances $Z_a$ and 
the correlation coefficients $f_{\rm v}$ and $f_\nu$. 
The common least-squares minimization (LSM) of the objective function 
is used to estimate the model parameters\footnote{
The LSM is computationally identical to the standard $\chi^2$-minimization,
but it can be used to evaluate the uncertainties of the fitting parameters
even in cases of correlated measurements. One must realize that the rebinning
used in the data reduction procedure introduces a correlation between
data point values (Levshakov et al. 2002)}.

The computational procedure itself consists of two steps: firstly a point 
in the parameter space ($N_0, U_0, \sigma_{\rm v}, \sigma_{\rm y}, 
Z_{\rm a})$  is chosen and then an optimal configuration 
of $\{v_i\}$ and $\{n_i\}$ for this parameter set is searched for. 
Correlation coefficients are considered as external parameters and 
remain fixed during the calculations.

\begin{figure*}
\vspace{0.0cm}
\hspace{0.0cm}\psfig{figure=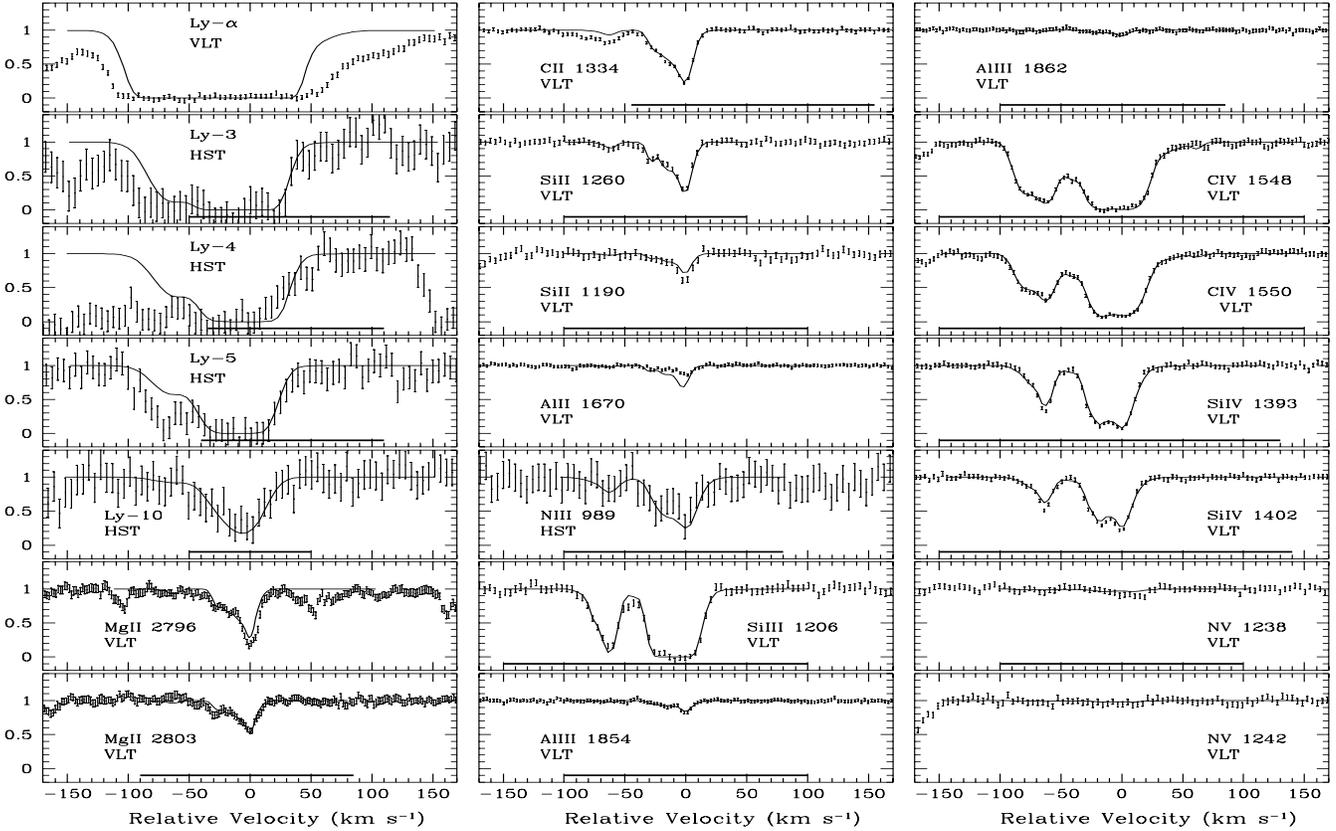,height=13.0cm,width=19.0cm}
\vspace{-2.0cm}
\caption[]{
Same as Fig.~1 but for the
$z_{\rm abs} \simeq 1.94$ system toward J2233--606.
The zero radial velocity is fixed at $z = 1.942616$. 
The corresponding physical parameters are listed in Table~1.
The normalized $\chi^2_{\rm min} = 1.08$ (the number of degrees of
freedom $\nu = 888$) 
}
\label{fig4}
\end{figure*}

The optimization of $\{v_i\}$ and $\{n_i\}$ is the most 
time-consuming part of the procedure and needs an effective algorithm 
to achieve a quick and stable convergence of the computations. 
In the MCI we use the simulated annealing with Tsallis acceptance 
rule (Xiang et al., 1997) and an adaptive annealing temperature choice. 
Namely, the annealing temperature $T_{\rm a}$ at iteration  $k+1$ 
is decreased according to following equation:
\begin{equation}
T^{k+1}_{\rm a} = \frac{ T^k_{\rm a} }{1 + \beta 
\frac{\Delta E_k}{E_k} r}\; ,
\label{eq:E8}
\end{equation}
where  $\Delta E_k/E_k$ is the relative energy
$(\equiv$ the sum of squares of the deviations, ${\cal S}$) 
variation on step $k$, $r$ is the acceptance rate, 
i.e. the ratio of the accepted trials to the total trial number, 
and $\beta$ is a constant of order 1.

The calculation of the uncertainty ranges 
for the fitting  parameters is 
in our method not so
straightforward as a simple inversion of 
the Hesse matrix since the
velocity and density distributions represent
additional degrees of freedom and widen, in general,  
the confidence intervals.
However,
these $\{v_i\}$ and $\{n_i\}$
distributions themselves are nuisance parameters and should be
`integrated out' when one computes 
the errors for the other parameters. 
To estimate the confidence levels, 
the following procedure can be applied: 
the values of  the physical parameters
in the vicinity of the global minimum of the objective function
are chosen at random
and then the optimal density and velocity distributions are computed. 
Assuming that the probability
of each parameter set can be linked to the derived ${\cal S}$ value 
[e.g. as $\exp(-{\cal S})$] we can estimate
from the obtained sample
the joint probability density function for parameters 
and hence calculate all necessary statistical moments.

It should be noted, however, that the reliable estimation 
of this multidimensional function requires
a very large sample which is  quite time consuming. 
In our case the exact estimation
of confidence levels is not very crucial taking into account 
the intrinsic uncertainties in atomic data (e.g. Savin 2000)
or unknown shape and intensity of the local
background ionizing radiation. 
Because of this
we restricted our samples to a few dozens of points  
and estimated the accuracy of the fitting parameters only approximately.

The recovered density and velocity 
patterns are not unique -- many configurations are possible with
comparable probability. 
But all these configurations have the same density-weighted
velocity distributions which actually determine the observed line shapes 
(see Paper~I). 
As already mentioned above, 
we represent these random fields by their values sampled at equally 
spaced intervals $\Delta x$. In order to compare the
calculated patterns we rearrange these values 
in such a way that the final configuration exhibits a lowest rate 
of entropy production: according to the Prigogine theorem 
(Prigogine 1967), this configuration has the minimal dissipation and, 
hence, is more stable and more probable. 
All necessary equations to calculate the entropy production
are presented in Appendix. 
We stress, however, that 
configurations produced on the base of such rearrangement
should in no case be considered as something final -- they represent only  
the (most) probable case of the density and velocity distributions
along the line of sight and are used here exclusively for 
illustrations.

\section{Results on individual systems}

\subsection{J2233--606, $z_{\rm abs} \simeq 1.92$}

The system at $z_{\rm abs} \simeq 1.92$ toward the quasar J2233--606 
has saturated hydrogen lines (from Ly-$\alpha$ up to Ly-8) 
and metal lines of 
\ion{C}{ii}, \ion{Si}{ii}, \ion{Si}{iii}, 
\ion{C}{iv},  and \ion{Si}{iv}. 
The results obtained with the MCI  
are presented in Table~1 and illustrated in
Figs.~1 and 2.
Parts of profiles included in the $\chi^2$ minimization 
are marked by horizontal lines at the panel bottoms in Fig.~1.
Profiles of the doublet \ion{N}{v} $\lambda1238$ \AA\, and 
\ion{N}{v} $\lambda1242$ \AA\, were 
calculated later using the obtained velocity and density
distribution and the metallicity derived from the fitting of 
\ion{N}{iii} $\lambda989$ \AA.
It is seen from Fig.~1 that most spectral features can be well 
represented assuming uniform
metallicities and a common HM UV background.
Fig.~2 demonstrates the distribution of the 
radial velocity and gas density (panels {\bf a} and {\bf b})
along the line
of sight (rearranged in accord with the principle 
of minimal entropy production rate).
The density distributions for the ions involved in the optimization  
are shown in panels {\bf c}-{\bf j}, 
whereas the density-weighted velocity distributions which
determine the shapes of the spectral lines are presented in
Fig.~3. 
This figure shows that the density-weighted 
velocity distributions 
for low ions \ion{C}{ii} and \ion{Si}{ii} are similar, but 
differ from those for high ions \ion{C}{iv} and \ion{Si}{iv}. 
These distributions easily explain why the lines of \ion{C}{ii} and 
\ion{Si}{ii} look very much alike and why their centers  are displaced 
by $\Delta v \simeq 4$ km~s$^{-1}$ (DP)
with respect to \ion{C}{iv} and \ion{Si}{iv}.

The study of this system by DP, 
who used the standard Voigt profile fitting, 
produced comparable column densities (albeit 20-50\% smaller). 
However, the metallicities obtained by DP for two main clouds 
(at $v$ = 0 km~s$^{-1}$ and $v = -43$ km~s$^{-1}$) differ 
nearly by two orders
of magnitude: [X/H] = --0.9 and --2.7, respectively 
(in our case [X/H] $\simeq -2.2$ for the whole system).
As shown in Paper~I, 
the Voigt fitting may in general yield correct column densities
when applied to unsaturated lines, but the mean $U$ and, hence, 
the ionization corrections may not be unambiguous. 
Therefore, 
the conclusion made by DP that the $z_{\rm abs} \simeq 1.92$ system contains
`a region of intense star-formation activity' may not be well justified 
since this result is model dependent.

\begin{figure}
\vspace{0.0cm}
\hspace{0.0cm}\psfig{figure=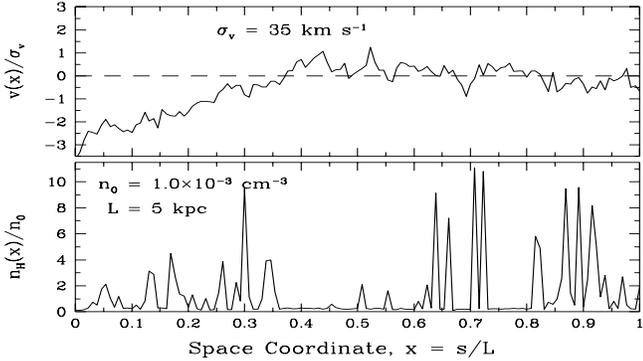,height=5.0cm,width=8.0cm}
\vspace{0.0cm}
\caption[]{Computed velocity (upper panel) and gas density (lower panel) 
distributions along the line of sight for the system at $z = 1.942616$
toward J2233--606. Shown are patterns rearranged according to the
principle of minimum entropy production rate (see text) }
\label{fig5}
\end{figure}

The values of the average gas density $n_0$ and kinetic temperature
$T_{\rm kin}$, and the
cloud thickness $L$
estimated in our model (see Table~1)
are typical for the Ly-$\alpha$ systems discussed in the literature 
(e.g., Giallongo \& Petitjean 1994; Viegas et al. 1999;
Prochaska \& Burles 1999; Chen et al., 1998; Chen et al. 2001). 
Low metallicity for the whole system ([X/H] $< -2.0$) 
and its dimension of 20~kpc imply that this system can originate 
in a galactic halo or in a large scale structure object.

\subsection{J2233--606, $z_{\rm abs} \simeq 1.94$}

This system exhibits a plenty of metal lines in different ionization stages. 
The metal profiles are not very complex
and extend over the velocity range from $-100$ km~s$^{-1}$ 
to 100 km~s$^{-1}$.
Results obtained with the MCI are presented in Table~1 and 
shown in Figs.~4 and 5. As in the previous system, most absorption
features can be well
described with uniform metallicities and a common HM spectrum.
The Ly-$\alpha$ profile is contaminated by the forest absorption
in the blue and red wings and therefore the Ly-$\alpha$ absorption feature
was not involved in the analysis.
The profiles of \ion{Mg}{ii}\,$\lambda2796$ \AA\, 
and \ion{Al}{ii}\,$\lambda1670$ \AA\, were computed later using the
derived velocity and density distributions. 
\ion{Mg}{ii}\,$\lambda2796$ \AA\, 
is contaminated by a telluric line and this explains
the difference between the computed and observed profiles. 
The synthetic and observed profiles of \ion{Al}{ii}\,$\lambda1670$ \AA\,
show much more pronounced discrepancy. 
Fractional ionisation curves for \ion{Al}{ii} and \ion{Al}{iii}
were computed with CLOUDY. 
These curves allowed us to fit the \ion{Al}{iii} doublet quite well 
with the Al abundance similar to that obtained for the other metals. 
However, when
the \ion{Al}{ii} profile was included
in the fitting, the Al metallicity differed by order 
of magnitude from the other metals. Besides
it was impossible to fit adequately the \ion{Al}{iii} doublet. 
Similar behaviour of Al was reported also by DP who noted
that `the recombination coefficients used to compute the 
aluminium ionisation equilibrium
[in CLOUDY] are probably questionable'.

Column densities derived by DP coincide well (within 15\%) with 
that obtained in our procedure except for the
saturated \ion{Si}{iii}\,$\lambda1206$ \AA\, line for which the Voigt 
fitting gave nearly 2 times lower value. 
The abundances estimated in DP scatter again from component 
to component, but nevertheless they conclude that 
`the gas in this system
is likely of quite high metallicity (larger than 0.1 solar)'. 

\begin{figure*}
\vspace{0.0cm}
\hspace{0.0cm}\psfig{figure=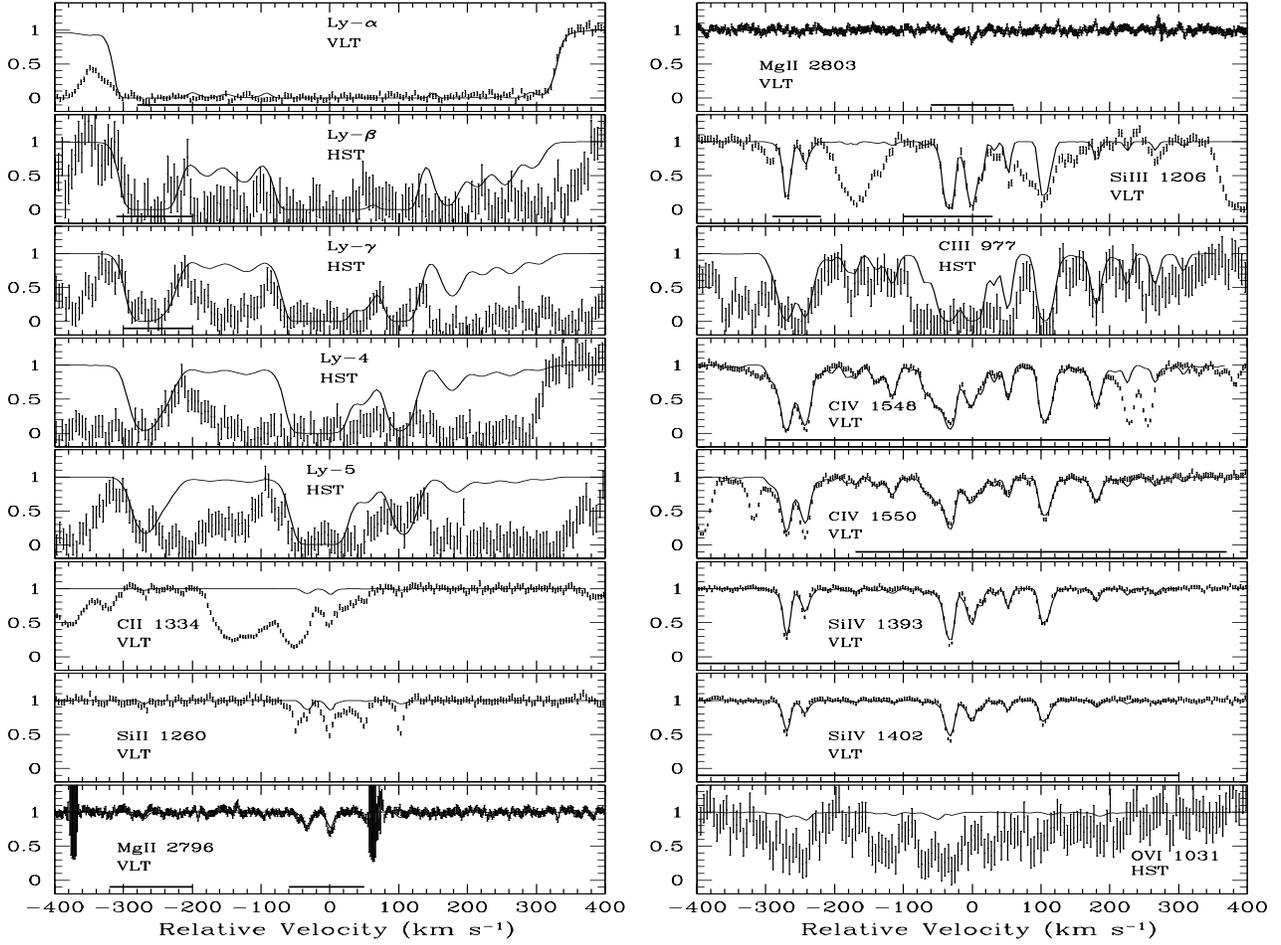,height=13.0cm,width=20.0cm}
\vspace{-0.5cm}
\caption[]{Same as Fig.~1 but for the
$z_{\rm abs} \simeq 1.87$ system toward J2233--606.
The zero radial velocity is fixed at $z = 1.87008$. 
The corresponding physical parameters are listed in Table~1.
The normalized $\chi^2_{\rm min} = 1.60$ (the number of degrees of
freedom $\nu = 1158$) 
}
\label{fig6}
\end{figure*}

Similar to the Voigt fitting,
the MCI also delivered for this system high metal abundancies:
one third solar for carbon and silicon and nearly 
two times lower for nitrogen, magnesium and aluminium. 
Taking into account this result and a compact dimension 
($\simeq 5$ kpc, see Table~1) 
of the absorbing region we come to the same conclusion as 
Prochaska \& Burles (1999) did:  the system at $z = 1.94$ can hardly  
be a large scale structure object (like a filament or a wall)
and should be related to a galactic system (may be a region of intense star
formation).

\subsection{J2233--606, $z_{\rm abs} \simeq 1.87$}

This is the most interesting system 
from the family of the absorbers at $z = 1.9$ 
toward J2233--606. The metal line profiles
show a rather complex structure extending over 
the velocity range of about 700 km~s$^{-1}$. 
Some of these profiles are severely blended that 
hampers the unique Voigt profile deconvolution (e.g. DP assumed 17 components
to describe metal profiles).

The MCI code turned out to be much more robust 
and was able to recover the self-consistent line profiles even under such
unfavourable conditions. The physical parameters which  the MCI delivered 
for the $z = 1.87$ system together with
the underlying velocity and density distributions  
are presented in Table~1 and in Figs.~6 and 7. It is seen from Fig.~6 
that like in the previous two systems all lines are well described 
with a single parameter set, uniform metallicities and a common  
HM UV background.
The blue wing of the Ly-$\alpha$ line is contaminated by the forest
absorption as is clearly seen from the Ly-$\beta$ and Ly-$\gamma$ profiles.
The synthetic
profile of the \ion{O}{vi}\,$\lambda1031$ \AA\, line 
was calculated later using  the derived best fitting
parameters and the oxygen abundance [O/H] = $-1.0$
(which is about 3 times over the other 
element abundances from this system).
Even with the increased abundance
the synthetic profile of \ion{O}{vi} is still much weaker than the observed
intensities. This discrepancy
rules out the ionization of \ion{O}{vi} by the adopted background
radiation. 
Taking into account that all other elements have been well
described with a given HM spectrum and that the collisional ionization
of oxygen
can hardly be effective at low densities ($n_{\rm H} \sim 10^{-5}$)
and temperatures of $\sim 25000$~K,
this result seems to favor the interpretation that 
the \ion{O}{vi} ion and the other ions do not arise 
in the same gas (Kirkman \& Tytler 1999; Reimers et al. 2001).

According to our results, the absorber at $z = 1.87$ 
could be a large size cloud 
with very high velocity dispersion. Its 
estimated linear size of 80 kpc is consistent 
with dimensions of extended gaseous envelopes observed around galaxies 
at $z < 1$. 
In these envelopes, \ion{Mg}{ii} absorption is the dominant observational 
signature at the distancies up to a few tens of kiloparsecs 
(Bergeron \& Boiss\'e 1991), 
whereas highly ionizied species like \ion{C}{iv} 
are observed at distances of at least 100 kpc from galactic centers
(Chen et al. 2001). 
Since the extended structure of the
same order of magnitude is observed at $z = 1.87$,  
we may conclude that this system arises 
in the external halo at large galactocentric distances.

\begin{figure}
\vspace{0.0cm}
\hspace{0.0cm}\psfig{figure=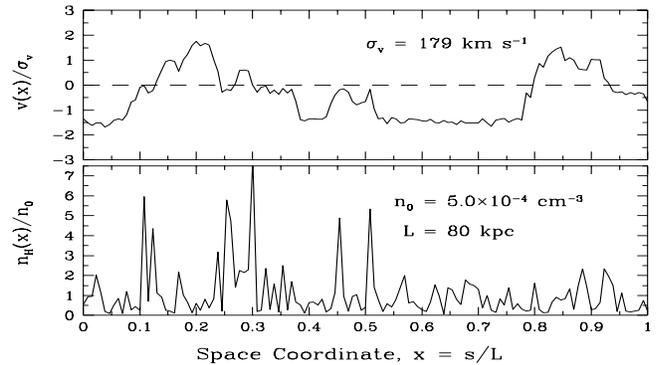,height=5.0cm,width=8.0cm}
\vspace{0.0cm}
\caption[]{Same as Fig.~5 but for the
$z_{\rm abs} \simeq 1.87$ system toward J2233--606 }
\label{fig7}
\end{figure}

\section{Discussion and conclusions}

A main goal of this work was to investigate the reliability of the physical
parameters and dynamical characteristics of the 
metal absorption systems obtained
by means of the standard Voigt fitting procedure and by the modified Monte
Carlo Inversion algorithm.
For comparison we used the recently obtained results on the Voigt fitting
analysis of three systems toward J2233--606 (D'Odorico \& Petitjean 2001).

We found that 
both approaches deliver similar
total column densities of unsaturated metal
lines. The saturated 
profiles may, however, be treated differently
(e.g., the \ion{Si}{iii} $\lambda1206$ \AA\, line
in the $z = 1.94$ system). 

We also found that metal abundances 
based on the Voigt deconvolution procedure
differ considerably from those obtained by the MCI.
For instance, instead of fluctuating metallicities
found in the absorbers toward J2233--606 by DP, 
the MCI shows that an assumption
of a homogeneous metal abundance for the whole system under study is quite
sufficient to represent all observed features.

New and principal results which can be obtained only with the MCI procedure
are the kinematic characteristics of the absorbers. We estimated 
selfconsistently for the first time the density and velocity dispersions
along the sightlines within the absorbers and calculated the total hydrogen
column and volumetric densities 
which gave us a direct measure of their linear sizes. 
The found dimensions of $L \simeq 5$ kpc to
$L \simeq 80$ kpc are in good agreement with measurements of the extended
gaseous envelopes around the nearby galaxes which were probed by the
\ion{Mg}{ii} absorption lines (Bergeron \& Boiss\'e 1991) and by
the \ion{C}{iv} lines (Chen et al. 2001).

A new issue obtained in the MCI analysis is 
a scaling relation. Namely,
we found that the linear size $L$ shows a positive correlation with 
the line-of-sight velocity dispersion $\sigma_{\rm v}$, i.e. 
the higher $L$, the
larger $\sigma_{\rm v}$ is observed (see Table~1).
Although our sample is still too small to carry out statistical analysis
of this correlation, 
the scaling tendency is of the same kind that can be expected
for virialized systems. The velocity dispersion is closely
related to the total mass of the system in a stationary state
(cf. the scaling law known as the {\it fundamental plane} for
elliptical galaxies). Taking into account that the scaling laws are
different for different types of objects (see, e.g., Fig.~2 in 
Mall\'en-Ornelas et al. 1999),
future statistical analysis may allow us to classify absorbers at
different redshifts.

It is also interesting to note another scaling relation:
we observe systematically higher metal abundance with decreasing $L$,
and vice verse, the higher $L$, the lower metallicity is deduced.
If $L$ reflects the linear size of a distant absorber, then we
may conclude that a compact absorber has, presumably, higher metal content
as compared with an extended one.

\begin{acknowledgements}
We thank our referee Prof. Reimers for his helpfull report.
S.A.L. and I.I.A. gratefully acknowledge the hospitality of the
Osservatorio Astronomico di Trieste and the National Astronomical
Observatory of Japan (Mitaka) where this work was performed.
We also thank Valentina D'Odorico for sharing with us the calibrated
VLT/UVES spectrum of J2233--606.
The work of S.A.L., I.I.A. and I.E.M. is partly supported by
the RFBR grant No. 00-02-16007.
\end{acknowledgements}

\begin{table*}
\centering
\caption{Physical parameters of the metal absorption systems toward J2233--606
derived by the MCI procedure}
\label{tab1}
\begin{tabular}{lccc}
\hline
\noalign{\smallskip}
Parameter & $z_{\rm abs}\simeq1.94$ & $z_{\rm abs}\simeq1.92$ & 
$z_{\rm abs}\simeq1.87$ \\
\noalign{\smallskip}
\hline
\noalign{\smallskip}
Mean ionization parameter, $U_0$ & $8.79_{-2}\,(\simeq 15$\%)$^\ast$ &
$2.99_{-2}\,(\simeq 20$\%) & 0.12 ($\simeq$ 10\%)  \\
Total H column density, $N_{\rm H}$, cm$^{-2}$ & $1.41_{19}\,(\simeq$ 15\%) &
$2.69_{20}\,(\simeq$ 10\%) & $1.23_{20}\,(\simeq$ 15\%) \\ 
Velocity dispersion, $\sigma_{\rm v}$, km~s$^{-1}$ & 35 ($\simeq$ 15\%) &
104 ($\simeq$ 20\%) & 179 ($\simeq$ 10\%) \\
Density dispersion, $\sigma_{\rm y}$ & 1.50 ($\simeq$ 15\%) &
2.31 ($\simeq$ 10\%) & 1.31 ($\simeq$ 10\%) \\
Chemical abundances$^\dagger$: & & &  \\ 
$Z_{\rm C}$ & $9.8_{-5}\, (\la$ 10\%) & $2.0_{-6}\, (\simeq$ 20\%) &
$8.7_{-6}\, (\la$ 10\%) \\
$Z_{\rm N}$ & $1.1_{-5}\, (\simeq$ 15\%) & $5.5_{-7}\, (\simeq$ 20\%) &
$\ldots$   \\
$Z_{\rm Mg}$ & $6.8_{-6}\, (\simeq$ 10\%) &$\ldots$& $2.0_{-6}\, (\la$ 10\%) \\ 
$Z_{\rm Al}$ & $5.1_{-7}\, (\simeq$ 15\%) &$\ldots$ &$\ldots$  \\
$Z_{\rm Si}$ & $1.0_{-5}\, (\la$ 10\%) & $2.5_{-7}\, (\simeq$ 20\%) &
$1.7_{-6}\, (\la$ 10\%) \\
$[Z_{\rm C}]$ & $-0.53$ &  $-2.21$ & $-1.58$  \\
$[Z_{\rm N}]$ & $-0.88$ &  $-2.17$ & $\ldots$   \\
$[Z_{\rm Mg}]$ & $-0.75$ & $\ldots$ & $-1.27$    \\
$[Z_{\rm Al}]$ & $-0.79$ & $\ldots$ & $\ldots$  \\
$[Z_{\rm Si}]$ & $-0.54$ & $-2.14$ & $-1.31$  \\
Column densities, cm$^{-2}$~: & & \\
$N$(\ion{H}{i}) & $(2.50\pm0.05)_{16}$ & $(2.2\pm0.25)_{17}$ &
$(2.65\pm0.28)_{16}$ \\
$N$(\ion{C}{ii}) & $(7.49\pm0.11)_{13}$ & $(1.80\pm0.20)_{13}$ &$\ldots$ \\
$N$(\ion{Mg}{ii}) & $(6.04\pm0.22)_{12}$ &$\ldots$  & $(1.96\pm0.21)_{12}$ \\
$N$(\ion{Si}{ii}) & $(8.79\pm0.25)_{12}$ & $(2.40\pm0.16)_{12}$ &
$(2.56\pm0.20)_{12}$  \\
$N$(\ion{C}{iii}) &$\ldots$ &$\ldots$ & $(5.36\pm0.25)_{14}$ \\
$N$(\ion{N}{iii}) & $(1.62\pm0.09)_{14}$ &$\ldots$ &$\ldots$ \\
$N$(\ion{Al}{iii}) & $(1.63\pm0.08)_{12}$ &$\ldots$ &$\ldots$ \\
$N$(\ion{Si}{iii}) & $(9.33\pm0.51)_{13}$ & $(2.90\pm0.18)_{13}$ &
$(4.94\pm0.25)_{13}$ \\
$N$(\ion{C}{iv}) & $(3.93\pm0.16)_{14}$ & $(4.78\pm0.31)_{13}$ &
$(3.66\pm0.13)_{14}$ \\
$N$(\ion{Si}{iv}) & $(5.58\pm0.35)_{13}$ & $(8.50\pm0.50)_{12}$ &
$(4.13\pm0.11)_{13}$ \\
$N$(\ion{N}{v}) & $< 3.54_{12}$ &$\ldots$  &$\ldots$    \\
Hydrogen number density, $n_0$, cm$^{-3}$ & $\simeq 1_{-3}$ & 
$\simeq 5_{-3}$ & $\simeq 5_{-4}$ \\
Mean kinetic temperature, K & $14_3$ & $25_3$ & $25_3$  \\
Minimum kinetic temperature, K & $11_3$ & $16_3$ & $18_3$  \\
Maximum kinetic temperature, K & $17_3$ & $40_3$ & $31.5_3$  \\
Linear size, $L$, kpc & $\simeq 5$ & $\simeq 20$ & $\simeq 77$  \\
Mass, $M^\ddagger$, in $M_\odot$ & $\simeq 1.4_6$ & $\simeq 4.5_8$ &
$\simeq 2.6_9$ \\
Mass, $M^\diamondsuit$, in $M_\odot$ & $\simeq 7.0_8$ & $\simeq 2.0_{10}$ &
$\simeq 3.0_{11}$ \\
\noalign{\smallskip}
\hline
\noalign{\smallskip}
\multicolumn{4}{l}{$^\ast$Values like `$a_x$' mean $a\times10^x$; 
shown in parenthesis are uncertainties at the $1\sigma$ c.l.}\\ 
\multicolumn{4}{l}{$^\dagger Z_{\rm X}$ = X/H,\, 
$[Z_{\rm X}] = \log (Z_{\rm X}) - \log (Z_{\rm X})_\odot$}\\
\multicolumn{4}{l}{$^\ddagger$Assuming spherical geometry;
$^\diamondsuit$Using the virial theorem}
\end{tabular}
\end{table*}

\appendix

\section{Minimum entropy production rate} 

\bigskip\noindent
General equation for the entropy production rate is given by 
(Landau \& Lifshits 1987):
\begin{eqnarray}
\lefteqn{
\frac{\partial S}{\partial t}\,= 
\int \frac{\kappa ({\rm grad}\, T)^2}{T^2}\, {\rm d}V\, + } \nonumber \\ 
\lefteqn{
\int \frac{\eta}{2T}\,\left(\frac{\partial v_i}{\partial x_k}\,+\,
\frac{\partial v_k}{\partial x_i}\,-\,\frac{2}{3}\delta_{ik}\,
\frac{\partial v_j}{\partial x_j}\right)^2\, {\rm d}V\,+ } \nonumber \\
\lefteqn{
\int \frac{\xi}{T}\,({\rm div}\, {\bf v})^2\, {\rm d}V\; .}
\label{eq:A1}
\end{eqnarray}
Here $\partial S/\partial t$ is the rate of the total entropy production
in the system,  
the velocity {\bf v} and temperature $T$ are functions of the coordinate
$(x,y,z)$, the volume element is denoted by d$V$,
the thermal conductivity, dynamic viscosity, and the second viscosity
coefficients -- by $\kappa$, $\eta$, and $\xi$, respectively.

The terms included in this equation account for only 
hydro- and thermodynamic processes since we assume the
ionization balance in each point of the region which
means that the radiative heating and cooling 
do not contribute to the entropy production. 
The second viscosity equals zero for dilute monoatomic gases 
(Chapman \& Cowling 1970)
so we omit it from further consideration.

We also assume that heat conductivity $\kappa$ and viscosity $\eta$ 
are dominated by turbulence, and hence the
Prandtl number is about unity (Monin \& Yaglom 1975):
\begin{equation} 
{\rm Pr} = \frac{\eta\, C_{\rm p}}{\kappa} \simeq 1\; ,  
\label{eq:A2}
\end{equation}
where $C_{\rm p}$ is a specific heat capacity at constant pressure.
The relation (\ref{eq:A2}) is valid for both the ionized and neutral gas. 

From observations, only one component $v_x$ (along the sightline)
of the velocity vector {\bf v} is known. 
Therefore we  are compelled
to neglect in (\ref{eq:A1}) all terms including derivatives 
other than $\partial v_x/\partial x$. 
The second right hand term in (\ref{eq:A1}) can be re-written in the form:
\begin{eqnarray}
\lefteqn{
\int \frac{\eta}{2T}\,\left(\frac{\partial v_i}{\partial x_k}\,+\,
\frac{\partial v_k}{\partial x_i}\,-\,\frac{2}{3}\delta_{ik}\,
\frac{\partial v_j}{\partial x_j}\right)^2\, {\rm d}V\,= } \nonumber \\
\lefteqn{
\int \frac{\eta}{2T}\,\left(\frac{\partial v_i}{\partial x_k}\,+\,
\frac{\partial v_k}{\partial x_i}\right)^2\, {\rm d}V\,-\,
\frac{4}{3}\int \frac{\eta}{2T}\,\left({\rm div}\, {\bf v}\right)^2\, {\rm d}V}
\label{eq:A1a}
\end{eqnarray}

After all these assumptions, we obtain the following simplified 1D form
for the entropy production rate 
\begin{eqnarray}
\lefteqn{
\frac{\partial \tilde{S}}{\partial t}\,= } \nonumber \\
\lefteqn{
\int \kappa \left[
\frac{1}{\tilde{T}^2}\,\left(\frac{{\rm d}\tilde{T}}{{\rm d}x}\right)^2\,+\, 
{\rm Pr}\,\frac{\sigma^2_{\rm v}}{C_{\rm p}\,T_0}\,
\frac{\cal A}{\tilde{T}}\,
\left(\frac{{\rm d}\tilde{v}_x}{{\rm d}x}\right)^2\right] {\rm d}x\; ,}
\label{eq:A3}
\end{eqnarray}
where $\tilde{v}_x = v_x/\sigma_{\rm v}$,
$\tilde{T} = T/T_0$, with $\sigma_{\rm v}$ and $T_0$ being
scales (characteristic values) 
for velocity and temperature inside the considered region, 
${\cal A}$ is a constant of about 1, 
and $\tilde{S}$ is the entropy per unit area.

For a monatomic ideal gas $C_{\rm p} = 5nR/2$ 
($n$ is the number of gram-moles and $R$ is the universal gas constant),
and if the gas is fully ionized then $\kappa \propto T^{5/2}$ 
(e.g. Lang 1999).

Given the values of $\kappa$, Pr, $T_0$, $C_{\rm p}$, and $\sigma_{\rm v}$,
we can rearrange the computational points 
$\{ v_1, \ldots, v_k \}$ and $\{ n_1, \ldots, n_k \}$ 
in such a manner, 
that minimum of (\ref{eq:A3}) will be achieved. 
Minimization of (\ref{eq:A3}) was carried out by means of combinatorial
simulated annealing technique (Press et al. 1992).
The obtained configuration of the density and velocity distributions 
should have the least dissipation
and therefore will exist longer than all others.


\begin{thebibliography}{}
\bibitem{}Bergeron J., \& Boiss\'e P. 1991, A\&A, 243, 344
\bibitem{}Chapman S., \& Cowling T. G. 1970, The Mathematical Theory of
Non-Uniform Gases, 3rd ed. (Cambridge Univ. Press: Cambridge)
\bibitem{}Chen H.-W., Lanzetta K. M., Webb J. K., \& Barcons X.
1998, ApJ, 498, 77
\bibitem{}Chen H.-W., Lanzetta K. M., \& Webb J. K.
2001, ApJ, 556, 158
\bibitem{}Cristiani S., \& D'Odorico V. 2000, AJ, 120, 1648
\bibitem{}D'Odorico V. \& Petitjean P. 2001, A\&A, 370, 729 (DP)
\bibitem{}Ferland G. J. 1997, Hazy, a brief introduction to
Cloudy 94.00 (http://www.pa.uky.edu/\~\,gary/cloudy)
\bibitem{}Giallongo E., \& Petitjean P. 1994, ApJ, 426, L61
\bibitem{}Haardt F., \& Madau P. 1996, ApJ, 461, 20
\bibitem{}Kirkman D. K., \& Tytler D. 1999, ApJ, 512, L5
\bibitem{}Landau L. D., \& Lifshitz E. M. 1987, Fluid Mechanics
(Butterworth-Heinemann)
\bibitem{}Lang K. R. 1999, Astrophysical Formulae, Vol. 1 (Springer-Verlag:
Berlin Heidelberg)
\bibitem{}Levshakov S. A., Takahara F., \& Agafonova I. I. 1999,
ApJ, 517, 609
\bibitem{}Levshakov S. A., Agafonova I. I. , \& Kegel W. H. 2000a,
A\&A, 355, L1  
\bibitem{}Levshakov S. A., Agafonova I. I. , \& Kegel W. H. 2000b,
A\&A, 360, 833 (Paper~I)
\bibitem{}Levshakov S. A., Dessauges-Zavadsky M., D'Odorico S., \& Molaro P.
2002, ApJ, 565, in press (astro-ph/0105529)
\bibitem{}Mall\'en-Ornelas G., Lilly S. J., Crampton D., \&
Schade D. 1999, ApJ, 518, L83
\bibitem{}Monin A. S., \& Yaglom A. M. 1975, Statistical Fluid Mechanics:
Mechanics of Turbulence, Vol. 2 (Cambridge: MIT)
\bibitem{}Press W. H., Teukolsky S. A., Vetterling W. T., \& 
Flannery B. P. 1992,
Numerical Recipes (Cambridge Univ. Press: Cambridge)
\bibitem{} Prigogine I. 1967, Thermodynamics of Irreversible
Processes (Wiley: New York)
\bibitem{}Prochaska J. K., \& Burles S. M. 1999, AJ, 117, 1957 
\bibitem{}Reimers D., Baade R., Hagen H.-J., \& Lopez S. 2001, A\&A, 374, 871
\bibitem{}Savaglio S. 1998, AJ, 116, 1055
\bibitem{}Savin D. W. 2000, ApJ, 533, 106
\bibitem{}Viegas S. M., Friaca A. C. S., \& Gruenwald R. 1999,
MNRAS, 309, 355
\bibitem{}Xiang Y., Sun D. Y., Fan W., \& Gong X. G. 1997, 
Phys. Lett. A, 233, 216
\end{thebibliography}
\end{document}